\documentclass[twocolumn,english,pra,showpacs,final,superscriptaddress,lengthcheck]{revtex4}
\usepackage{ae,aecompl}
\usepackage[T1]{fontenc}
\usepackage[latin9]{inputenc}
\usepackage{amsmath}
\usepackage{graphicx}
\usepackage{amssymb}

\makeatletter
\@ifundefined{textcolor}{}
{%
 \definecolor{BLACK}{gray}{0}
 \definecolor{WHITE}{gray}{1}
 \definecolor{RED}{rgb}{1,0,0}
 \definecolor{GREEN}{rgb}{0,1,0}
 \definecolor{BLUE}{rgb}{0,0,1}
 \definecolor{CYAN}{cmyk}{1,0,0,0}
 \definecolor{MAGENTA}{cmyk}{0,1,0,0}
 \definecolor{YELLOW}{cmyk}{0,0,1,0}
 }

\@ifundefined{definecolor}
 {\@ifundefined{definecolor}
 {\usepackage{color}}{}
}{}
\usepackage{epsfig}

\usepackage{amsfonts}
\usepackage{float}

\makeatother

\usepackage{babel}

\makeatother

\usepackage{babel}

\makeatother

\usepackage{babel}

\makeatother

\makeatother

\usepackage{babel}

\makeatother

\usepackage{babel}

\begin{document}

\title{Spin chains for robust state transfer: Modified boundary couplings
vs. completely engineered chains}

\author{Analia Zwick}

\affiliation{Fakultät Physik, Technische Universität Dortmund, D-44221 Dortmund,
Germany.}

\affiliation{Facultad de Matemática, Astronom\'{i}a y F\'{i}sica and Instituto
de F\'{i}sica Enrique Gaviola, Universidad Nacional de Córdoba, 5000
Córdoba, Argentina.}

\author{Gonzalo A. Álvarez}

\affiliation{Fakultät Physik, Technische Universität Dortmund, D-44221 Dortmund,
Germany.}

\author{Joachim Stolze}

\affiliation{Fakultät Physik, Technische Universität Dortmund, D-44221 Dortmund,
Germany.}

\author{Omar Osenda}

\affiliation{Facultad de Matemática, Astronom\'{i}a y F\'{i}sica and Instituto
de F\'{i}sica Enrique Gaviola, Universidad Nacional de Córdoba, 5000
Córdoba, Argentina.}


\begin{abstract}
Quantum state transfer in the presence of static disorder and noise
is one of the main challenges in building quantum computers. We compare
the quantum state transfer properties for two classes of qubit chains
under the influence of static disorder. In fully engineered chains
all nearest-neighbor couplings are tuned in such a way that a single-qubit
state can be transferred perfectly between the ends of the chain,
while in chains with modified boundaries only the two couplings between
the transmitting and receiving qubits and the remainder of the chain
can be optimized. We study how the disorder in the couplings affects
the state transfer fidelity depending on the disorder model and strength
as well as the chain type and length. We show that the desired level
of fidelity and transfer time are important factors in designing a
chain. In particular we demonstrate that transfer efficiency comparable
or better than that of the most robust engineered systems can also
be reached in chains with modified boundaries without the demanding
engineering of a large number of couplings. 
\end{abstract}
\maketitle

\section{Introduction}

One of the main challenges on the road to practical quantum computers
is the reliable transfer of quantum information between quantum gates
\cite{Divicenzo}. The main source of problems is the vulnerability
of quantum systems to perturbations due either to manufacturing imperfections
or to interactions with the environment. Overcoming (or avoiding)
these problems has motivated an intensive search for systems able
to transfer information with high quality while at the same time requiring
minimal control in order to avoid the introduction of errors \cite{Bose2008,Kay2010}.
The problem of state transfer has received a lot of attention in the
last decade in the context of quantum-information processing; nevertheless,
an early antecedent can be found in the work of Shore and co-workers,
see, for example, the paper of Cook and Shore \cite{Cook1979}.

Spin chains are a promising class of systems to serve as reliable
quantum communication channels \cite{Bose2003,Linden2004,Burgarth2005,Burgarth2007,Bose2008,Kay2010,Lukin2011,Zwick2011b,Bose2011}.
Perfect state transfer (PST) without any dynamical control can be
achieved by an infinity of engineered spin-spin coupling configurations
\cite{Albanese2004,Christandl2004,Stolze2005,Kay2006,Wang2011,Zwick2011b,Nikolopoulos2004,Kostak2007}
for a spin chain of given length. Regrettably this amazing transfer
fidelity comes at a high price in terms of the accuracy required to
design each interaction to avoid the loss of information \cite{DeChiara2005,Kay2010,Zwick2011b,Damico2011}.

In order to assess the reliability of these systems as realistic channels
for information transfer it is therefore essential to study the influence
of imperfections. Indeed, we have explored the robustness of some
PST channels against static perturbations \cite{Zwick2011b}, finding
that the quality of transfer is often strongly impaired by perturbations.
Therefore a question emerges: Is it really necessary to optimize every
single interaction in a chain? Can we find simpler systems showing
good transfer under perturbations?

In this work we focus on the behavior of essentially homogeneous chains
where only the first and last couplings can be adjusted. We show that
under perturbations these chains can achieve an optimized state transfer
(OST) comparable to or even better than that of fully engineered PST
systems. Two interesting regimes for transmission can be observed
when the boundary couplings are varied; for unperturbed chains these
regimes have already been studied recently \cite{Wojcik2005,Apollaro2010,Feldman2010,Zwick2011,Lukin2011,Apollaro2011,Bose2011}.
Favorable values for the speed and fidelity of transmission were observed
\emph{(i)} for an optimized (length-dependent) value of the boundary
couplings which renders quantum state transfer approximately dispersionless
and \emph{(ii)} in the limit of weak boundary couplings. For both
regimes we study the robustness against perturbations, demonstrating
that transfer efficiency comparable or better than that of the most
robust PST systems can be reached without the demanding engineering
of a large number of couplings.

\section{Spin chains as state transfer channels}

We consider a spin-$\frac{1}{2}$ chain with XX interactions between
nearest neighbors, described by the Hamiltonian\begin{equation}
H=\frac{1}{2}\sum_{i=1}^{N-1}J_{i}\left(\sigma_{i}^{x}\sigma_{i+1}^{x}+\sigma_{i}^{y}\sigma_{i+1}^{y}\right),\label{eq:hamiltonian}\end{equation}
 where $\sigma_{i}^{x,y}$ are the Pauli matrices, $N$ is the chain
length, and $J_{i}>0$ is the exchange interaction coupling. We assume
the mirror symmetry $J_{i}=J_{N-i}$, which is essential for PST \cite{Bose2008,Kay2010}.
These spin chains may be modeled with flux qubits \cite{Romito2005,Plenio2007,Strauch2008},
quantum dots \cite{Nikolopoulos2004,Lambropoulos2004,Lambropoulos2006},
atoms in optical lattices \cite{Plenio2007-1,Zoller2003,Lewenstein2007,Lukin2003},
and nitrogen vacancy centers in diamond \cite{Lukin2011}.

The goal is to transmit a quantum state $\left|\psi_{0}\right\rangle $
initially stored on the first spin ($i=1$) to the last spin of the
chain ($i=N$). $\left|\psi_{0}\right\rangle $ is an arbitrary normalized
superposition of the spin down ($\left|0\right\rangle $) and spin
up ($\left|1\right\rangle $) states of the first spin, with the remaining
spins of the chain initialized in a spin down state. Note that more
general initial states can be treated without much additional effort,
since the Hamiltonian (\ref{eq:hamiltonian}) is equivalent to one
of non-interacting fermions. The Hamiltonian (\ref{eq:hamiltonian})
conserves the number of up spins, $[H,\Sigma_{i}\sigma_{i}^{z}]=0$.
Therefore the component $|\mathbf{0}\rangle=|00...0\rangle$ of the
initial state is an eigenstate of $H$ and only the component $\mathbf{|1}\rangle=\mathbf{|}1_{1}0......0\rangle$
evolves within the one excitation subspace spanned by the basis states
$\mathbf{|i}\rangle=\mathbf{|}0...01_{i}0...0\rangle$. To evaluate
how well an unknown initial state is transmitted, we use the transmission
fidelity, averaged over all possible $\left|\psi_{0}\right\rangle $
from the Bloch sphere (see for details, Ref. \cite{Bose2003})\begin{eqnarray}
F & = & \frac{f_{N}}{3}\cos\gamma+\frac{f_{N}^{2}}{6}+\frac{1}{2},\label{eq:Averaged-Fidelity}\end{eqnarray}
 where $f_{N}^{2}=\left|\langle\mathbf{N}|e^{-\frac{iHt}{\hslash}}|\mathbf{1}\rangle\right|^{2}$
is the fidelity of transfer between states $|\mathbf{1}\rangle$ and
$|\mathbf{N}\rangle$ and $\gamma=\arg\left|f_{N}(t)\right|$. Because
the phase $\gamma$ can be controlled by an external field once the
state is transferred, we consider $\cos\gamma=1$. By the symmetries
of the system, this fidelity can be expressed in terms of the single-excitation
energies $E_{k}$ and the eigenvectors $|\Psi_{k}\rangle$ of $H$,
in the following way\begin{equation}
f_{N}=\left|\sum_{k,s}(-1)^{k+s}P_{k,1}P_{s,1}e^{-i(E_{k}-E_{s})t}\right|\label{eq:fn}\end{equation}
 where $P_{k,1}=a_{k,1}^{2}$ are the eigenvector probabilities on
the first site of the chain, since $|\mathbf{i}\rangle=\sum a_{k,i}|\Psi_{k}\rangle$.
PST channels are distinguished by commensurate energies $E_{k}$,
that is, all transition frequencies share a common divisor to make
$f_{N}=1$ in Eq. (\ref{eq:fn}) at a suitable PST time $\tau_{_{PST}}$
\cite{Christandl2004,Stolze2005}. This condition is obtained by suitably
modulating the spin-spin couplings $J_{i}$ \cite{Kay2006,Stolze2005,Zwick2011b}.

A long unmodulated homogeneous spin channel, $J_{i}=J$ $\forall i$,
cannot transfer a state perfectly, since due to the dispersive quantum
dynamics the transfer fidelity decreases with the number of spins
in the channel \cite{Bose2003}. In fact, rigorous PST in a homogeneous
chain is possible only for $N\leq3$ \cite{Christandl2004,Christandlpra}.
However, transfer can be noticeably improved just by lowering the
couplings of the spins at the ends of the channel.

We consider the two surface spins $i=1$ and $N$ interacting with
the inner spins with \textbf{$J_{1}=J_{N-1}=\alpha J$} while the
remaining spins compose a homogeneous chain with $J_{i}=J$. We call
this Hamiltonian $H^{\alpha}$, where $\alpha\in(0,1]$ is a control
parameter. This system has already been studied in Refs. \cite{Wojcik2005,Feldman2010,Zwick2011,Apollaro2011,Lukin2011}.

Two regimes for $\alpha$ can be used for OST: \emph{(i)} the \textit{optimal-coupling
regime} ($\alpha=\alpha_{opt}\sim N^{-\frac{1}{6}}$) possessing an
almost equidistant spectrum $E_{k}$ in the middle of the energy band,
resulting in a quasi dispersionless fast transfer with high fidelity
\cite{Zwick2011,Apollaro2011,Bose2011}; and \emph{(ii)} \textit{the
weak-coupling regime} ($\alpha\ll1)$. In that regime the transmitted
state appears and then reappears roughly periodically at the receiving
end of the chain. \emph{A}\textit{lmost perfect transfer} is achieved
with the first arrival due to the fact that only very few eigenstates
from the center of the energy band are involved, which are highly
localized at the boundaries of the chain \cite{Wojcik2005,Zwick2011,Lukin2011}.

The characteristic features of the two regimes just mentioned were
also observed to be essential for the robustness of PST spin-chain
channels against perturbations \cite{Zwick2011b}. The most robust
systems either showed an equidistant (linear) energy spectrum generating
the analog of dispersionless wave packet transfer or a large density
of states in the center of the band with the corresponding eigenstates
localized at the boundary sites of the chain and thus dominating the
end-to-end transfer \cite{Zwick2011b}. A class of PST systems is
characterized by a power-law spectrum\textbf{ }$E_{k}=sgn(k)|k|^{m},$
where $k=-\frac{{N-1}}{2},...,\frac{{N-1}}{2}$ and the exponent $m$
is a positive integer. We specifically address the linear case, $m=1$,
and call the corresponding Hamiltonian $H^{lin}$ \cite{Christandl2004},
and similarly for the quadratic case, $m=2$, with Hamiltonian $H^{quad}$
\cite{Zwick2011b}.

The OST system described by $H^{\alpha}$, requiring control of only
two boundary couplings, would obviously be simpler to implement than
the PST systems requiring engineering of all couplings along the chain.
In the following we compare the transmission performance of OST and
PST systems under the influence of disordered couplings in the \textit{channel}
assuming perfect control of the boundary couplings. We make the same
assumption for the engineered chains.

\begin{figure}
\includegraphics[width=1\columnwidth,height=0.7\columnwidth]{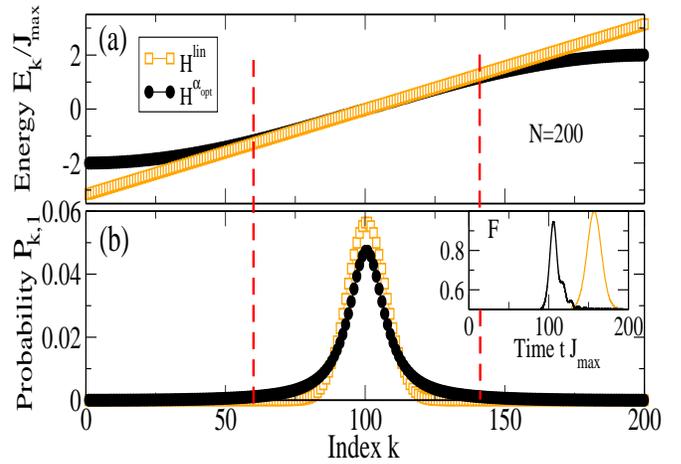}

\caption{\label{fig1}(Color online) Properties of the $H^{\alpha_{opt}}$
system (black solid dots) and the $H^{lin}$ system (orange open squares)
for a chain length $N=200$. (a) Eigenenergies $E_{k}$. (b) Probabilities
$P_{k,1}$ of the initial state $\left|\psi_{0}\right\rangle =|\mathbf{1}\rangle$.
The dashed vertical lines show the dominant energy eigenstates $|k\rangle$
that contribute to the state transfer. $P_{k,1}^{\alpha_{opt}}$ is
Lorentzian and $P_{k,1}^{lin}$ Gaussian. Inset: Evolution of the
averaged fidelity of the state transfer. }

\end{figure}

Static disorder in the couplings within the transfer channel is described
by $J_{i}\rightarrow J_{i}+\Delta J_{i}$ $(i=2,...,N-2)$, with $\Delta J_{i}$
being a random variable. We consider two possible coupling disorder
models:\textbf{ (a)} \textit{relative static disorder (RSD)} \cite{Zwick2011b,DeChiara2005,Petrosyan2010}\textit{,}
where each coupling is allowed to fluctuate by a certain fraction
of its ideal size, $\Delta J_{i}=J_{i}\delta_{i}$, and \textbf{(b)}
\textit{absolute static disorder (ASD) ,} where all couplings may
fluctuate within a certain fixed range which we measure in terms of
$J_{max}=\max J_{i}$: $\Delta J_{i}=J_{max}\delta_{i}$ \cite{Damico2011}.
Each $\delta_{i}$ is an independent and uniformly distributed random
variable in the interval $\left[-\varepsilon_{J},\varepsilon_{J}\right]$.
$\varepsilon_{J}>0$ characterizes the strength of the disorder. The
two coupling disorder models are equivalent for the OST systems since
all couplings are equal there. However, in the fully engineered PST
systems $J_{max}-J_{min}$ depends on the type of system and tends
to increase with $N$ so that absolute disorder is expected to be
more damaging than the relative one in these systems. The relevant
kind of disorder depends on the particular experimental method used
to engineer the spin chains.

\subsection{\textit{\emph{Optimal coupling regime}}}

When $\alpha=\alpha_{opt}$ in $H^{\alpha}$ the spectrum is linear
in the middle of the energy band {[}Fig. \ref{fig1}(a){]}. The probability
$P_{k,1}$ of the $k$th energy eigenstate to participate in the state
transfer is shown in Fig. \ref{fig1}(b) as a function of $k$ for
$N=200$. The linear part of the spectrum evidently dominates the
dynamics. Also shown in the figure are the corresponding quantities
for the linear PST chain. The obvious similarities between these two
systems suggest a comparison of their properties in the perturbed
case, which is discussed below. The inset in Fig. \ref{fig1}(b) shows
the averaged transfer fidelity of the unperturbed linear PST and $\alpha_{opt}$
systems, as a function of time. The maximum fidelity of the $\alpha_{opt}$
system is clearly smaller than unity, and it decreases with each revival
of the signal. However, the transfer time $\tau$ of the $\alpha_{opt}$
system is shorter: $\tau^{lin}=\frac{\pi N}{4J_{max}}$ \cite{Christandl2004,Zwick2011b}
and $\tau^{\alpha_{opt}}\sim\frac{N}{2J_{max}}$ \cite{Zwick2011};
hence $\tau^{lin}\sim\frac{\pi}{2}\tau^{\alpha_{opt}}$.

\begin{figure*}
\includegraphics[width=1.6\columnwidth,height=1\columnwidth]{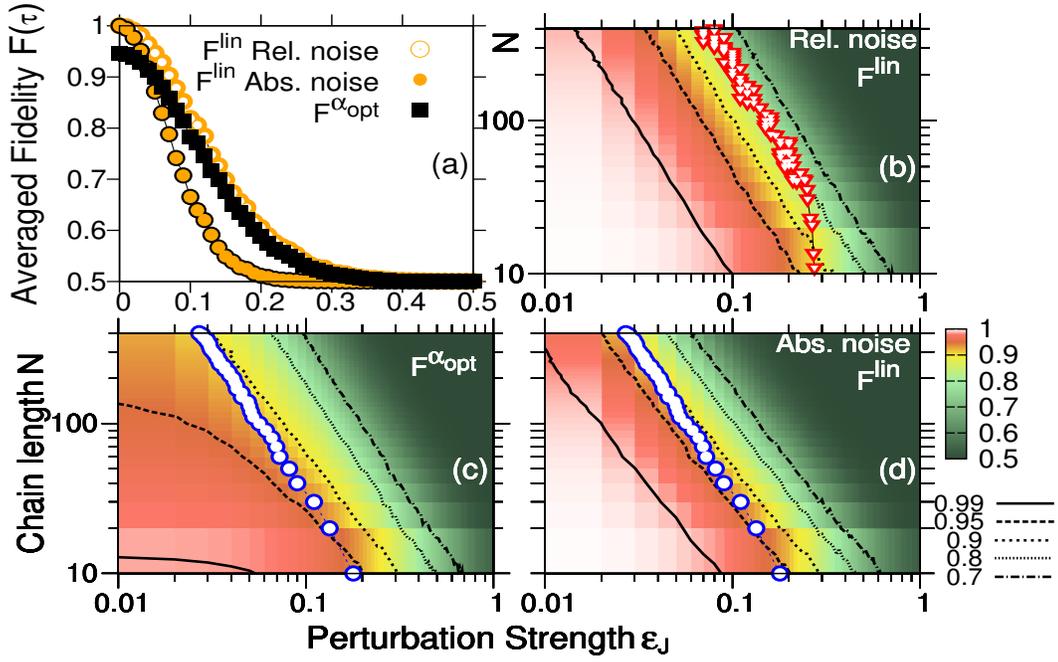}

\caption{\label{fig2}(Color online) Averaged fidelity at time $\tau$ as a
function of the perturbation strength $\varepsilon_{J}$ and the chain
length $N$ for $H^{lin}$ and $H^{\alpha_{opt}}$ systems. Relative
and absolute static disorder are considered. (a) $F^{lin}$ with relative
disorder (open circles) and absolute disorder (orange circles) and
$F^{\alpha_{opt}}$ (black squares) for both kinds of disorder when
N=200. (b) $F^{lin}$ with relative disorder . The open triangles
indicate when $F^{lin}=F^{\alpha_{0}}$ shown in Fig 3(c). To the
left of the symbols $F^{lin}>F_{odd}^{\alpha_{0}}$ (the difference
being small, however), while to the right $F^{\alpha_{0}}>F^{lin}$.
(c) $F^{\alpha_{opt}}$ with both kinds of disorder and (d) $F^{lin}$
with absolute disorder . The open circles indicate when $F^{lin}=F^{\alpha_{opt}}$.
To the left of the symbols $F^{lin}>F^{\alpha_{opt}}$ and to the
right $F^{\alpha_{opt}}>F^{lin}$. }

\end{figure*}

The main results of the comparison between the linear PST and $\alpha_{opt}$
systems are shown in Fig. \ref{fig2}. Figure \ref{fig2}(a) shows
the fidelity at time $\tau$ given by the transfer time of the unperturbed
case. The transfer fidelity is averaged over the Bloch sphere, as
well as over the disorder, for $N=200$, as a function of the disorder
strength $\varepsilon_{J}$. As expected the linear PST chain with
RSD always performs better than that with ASD. For vanishing disorder
strength the linear PST chain yields unit fidelity, which the boundary-controlled
chain does not, since its energy spectrum is only approximately, but
not strictly linear. The linear PST system with RSD has fidelity higher
than that of the boundary controlled system for all $\varepsilon_{J}$,
but for $\varepsilon_{J}\gtrsim0.1$ (where the fidelity is already
rather low) the difference in fidelity between the two systems becomes
insignificant. However, with ASD , there is a finite perturbation
strength ($\varepsilon_{J}\approx0.05$) where the $\alpha_{opt}$
system becomes better than the linear PST system. Hence, if fidelity
very close to unity is desired, complete engineering of the couplings
and very good disorder protection are mandatory. However, if only
moderate fidelity is needed (or possible, due to high disorder level)
a boundary-controlled system might do.

In order to see how the transfer properties depend on the chain length
we show in Figs. \ref{fig2}(b)-\ref{fig2}(d) the average fidelity
for each of the three systems as a contour and color plot in the $(\varepsilon_{J},N)$
plane. The contour lines are straight lines (representing power laws)
in most cases, with deviations for the boundary-controlled system
at weak disorder. The open circles in Figs. \ref{fig2}(c) and d indicate
where the fidelity of the boundary-controlled chain is equal to that
of the linear PST chain with ASD; to the right of them the boundary-controlled
chain has higher fidelity.

The results above already indicate that there is no simple general
answer to the question whether fully engineered or boundary-controlled
spin chains provide better quantum state transfer properties in the
presence of disorder. The static disorder model, strength and chain
length all are important factors in answering that question. We arrive
at similar conclusions in our next example.\\
\begin{figure*}
\includegraphics[width=1.6\columnwidth,height=1\columnwidth]{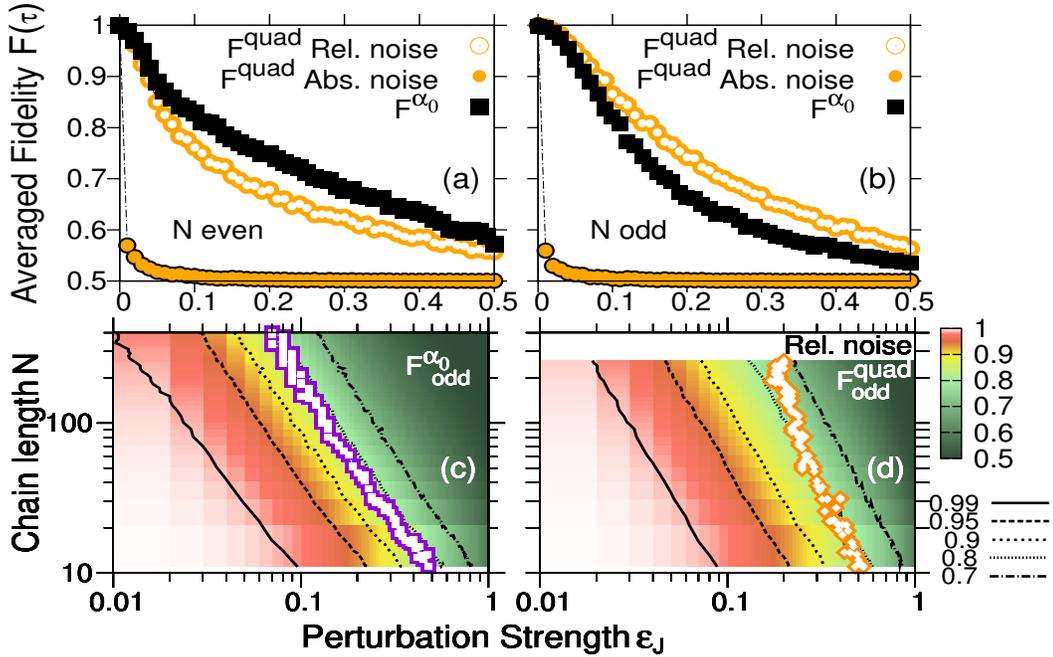}

\caption{\label{fig3} (Color online) Averaged fidelity at time $\tau$ as
a function of the perturbation strength $\varepsilon_{J}$ and of
the chain length $N$ for $H^{quad}$ and $H^{\alpha_{0}}$ systems
when relative and absolute disorder are considered. (a) $F^{quad}$
with relative disorder (open circles) and absolute disorder (orange
circles) and $F^{\alpha_{0}}$ (black squares) for both kinds of disorder
when $N=200$. (b) Same as panel (a) for $N=201$. (c) $F^{\alpha_{0}}$
with both kinds of disorder for odd $N$. The open squares show when
$F_{odd}^{\alpha_{0}}=F_{even}^{\alpha_{0}}$(not shown), where $F_{odd}^{\alpha_{0}}>F_{even}^{\alpha_{0}}$
to the left of the symbols. (d) $F_{odd}^{quad}$ with relative disorder.
The open diamonds indicate when $F_{even}^{\alpha_{0}}=F_{odd}^{quad}$,
where $F_{odd}^{quad}>F_{even}^{\alpha_{0}}$ to the left of the symbols.}

\end{figure*}

\subsection{\textit{\emph{Weak Coupling Regime}}}

When the boundary spins are only weakly coupled to the channel, i.e.,
$\alpha J_{max}=\alpha_{0}J_{max}\ll\frac{1}{\sqrt{N}}$ in $H^{\alpha}$,
an almost perfect state transfer, $F\approx1-$\textbf{$\mathcal{O}$}$(\alpha^{2}J_{max}^{2}N)$,
is achieved (for details, see Ref. \cite{Wojcik2005}). In this region,
the parity of $N$ is relevant. This can be understood by studying
the spectral properties of the {}``channel'' of $N-2$ spins connecting
the transmitting and receiving qubits. For odd (even) $N$ the dynamics
of the channel is dominated by two (three) states situated symmetrically
about the center of the energy spectrum \cite{Wojcik2005}. The energy
differences between these dominant levels determine the transfer time
which is obtained as $\tau{}_{even}^{\alpha_{0}}\sim\frac{\pi}{2\alpha^{2}J_{max}}$
and $\tau_{odd}^{\alpha_{0}}\sim\frac{\pi\sqrt{N}}{2\alpha J_{max}}$
\cite{Wojcik2005}. Since $\tau$ is $N$ independent for even $N$
and $\alpha_{0}J_{max}<\frac{1}{\sqrt{N}}$, the transfer is faster
for odd $N$. Very similar properties of the energy eigenstates which
dominate the state transfer are found in the fully engineered (PST)
chain with odd $N$ and a quadratic energy spectrum, which make it
the most robust PST system for relative disorder \cite{Zwick2011b}.
We therefore compare this system to the boundary-controlled chain
at weak coupling. We find that the transfer time of the quadratic
PST chain is $\tau^{quad}\sim\frac{\pi N^{2}}{8J_{max}}$ which is
longer than $\tau_{odd}^{\alpha_{0}}$ for $\alpha\gtrsim\frac{4}{N^{3/2}}$
for reasonably large $N$.

Figures \ref{fig3}(a) and (b) show the averaged fidelities for $N=200$
and $N=201$, respectively, for the quadratic PST system and the weak-coupling
boundary-controlled system, at time $\tau$ determined by the unperturbed
cases and for $\alpha=0.01$. Again, as in the linear case, absolute
disorder is much more detrimental than relative disorder. This is
connected to the fact that the maximum and minimum couplings in the
chain may differ by orders of magnitude, with the small couplings
always close to the ends of the chain \cite{Zwick2011b}. Consequently
a fluctuation of a given absolute size may completely spoil the state
transport when it affects one of the small couplings close to the
boundary. For the boundary-controlled system the two kinds of disorder
are again equal by definition. Therefore, for absolute disorder the
weak-coupling OST system performs always better than the quadratic
PST system. For relative disorder the parity of $N$ matters. The
fidelity of the boundary-controlled system is similar or higher (lower)
than that of the PST system when $N$ is even (odd). Figures \ref{fig3}(c)
and \ref{fig3}(d) show the fidelity as a contour and color plot in
the $(\varepsilon_{J},N)$ plane for $\alpha=0.01$ and odd $N$.
The contour lines are again power laws. The open symbols in Fig. \ref{fig3}(c)
(squares) indicate where the fidelities for odd and even weak-coupling
boundary-controlled systems are equal. To the left of the symbols
the fidelity is higher for odd $N$. The open symbols (diamonds) in
Fig. \ref{fig3}(d) indicate where the fidelities for odd quadratic
PST systems (with relative disorder) and for even weak-coupling boundary-controlled
systems are equal. To the left of the symbols the fidelity is higher
for the quadratic PST system, but for small perturbation strength
differences between the two systems are quite small.

We want to remark that if an actual implementation were to be used,
the faulty couplings of the chain could be tested following the recipe
given in Ref. \cite{Wie2010}, which allows the coupling strength
estimation of a $XX$ spin chain with an external magnetic field applied
to it. In this case the best possible time to remove the state from
the chain can be obtained from the numerical integration of the Schrödinger
equation, just looking for the smallest time when the fidelity is
near 1. In case the indirect Hamiltonian tomography \cite{Burgarth2011}
turns out to be too expensive or cumbersome to perform, the best time
to remove the state from the chain is the design time, {\em i.e.},
the time $\tau$ when the fidelity of the {}``nonfaulty chain'',
the one that was intended to be implemented, achieves its best performance.

On the other hand, a detailed analysis of the statistics of the fidelity
as a function of time is lacking; so far most studies focus on its
average over realizations of the noise. For a particular class of
engineered chains \cite{Christandl2004}, De Chiara \emph{et al.}
\cite{DeChiara2005} have shown that the time signal of the fidelity
becomes fractal. In this sense, it is difficult to assess how much
information is lost because of a bad timing for the readout of the
state at the receiving end of the channel.

\section{Summary and conclusions}

\begin{figure}
\includegraphics[width=0.9\columnwidth]{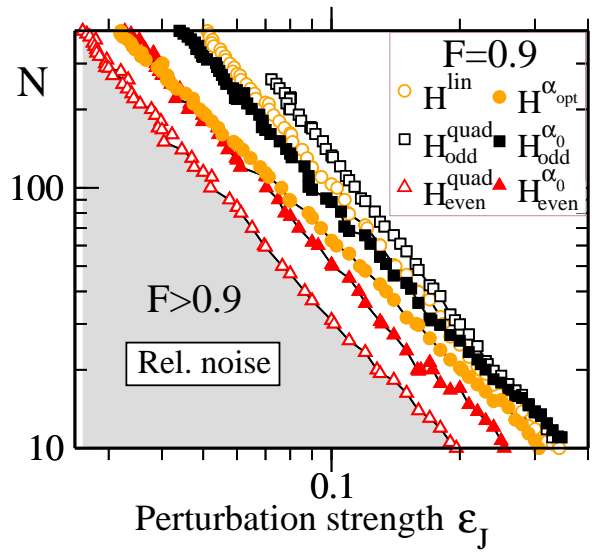}\caption{\label{fig4}(Color online) Contour lines of the averaged transfer
fidelity $F=0.9$ for fully-engineered PST systems (closed symbols)
and boundary-controlled $\alpha-$OST systems. To the left of the
symbols the transfer fidelity $F>0.9$ for every system.}

\end{figure}

For relative disorder, Fig. \ref{fig4} shows a comparison between
all of the systems considered here, linear PST and boundary-controlled
with optimal $\alpha_{opt}$ as well as quadratic PST and weak-coupling
$(\alpha=0.01)$ boundary-controlled, for both even and odd lengths.
For each system the figure shows the line in the $(\varepsilon_{J},N)$
plane where $F=0.9$. Open symbols denote PST systems; closed symbols
correspond to boundary-controlled systems. To the left of the symbols
the transfer fidelity of each system is $F>0.9$. It is interesting
to note that the lines for the three boundary-controlled systems lie
next to each other (at least for long chains), while one of the PST
systems (quadratic, even) lies clearly below (performs less well)
and the other two PST systems lie slightly above. This situation changes,
however, for different levels of fidelity. For example, the $H_{odd}^{\alpha_{0}}$
system outperforms $H^{lin}$ in the region to the right of the crossover
marked by the open triangles in Fig. \ref{fig2}(b). Note that to
the left of that crossover the fidelities of the two systems differ
only by up to 4\%. On the other hand, to the right of the crossover
displayed in Fig. \ref{fig3}(d), $H_{even}^{\alpha_{0}}$ is the
best choice.%

For absolute disorder, there is almost always a boundary-controlled
system with fidelity larger than that of the PST systems. Only for
very small perturbation strength can PST systems be better than OST
systems, but the fidelities are similar.

Considering only the PST systems, $H_{odd}^{quad}$ performs better
than $H^{lin}$ for relative disorder with similar transfer fidelity
for small perturbations. Conversely, $H^{lin}$ is drastically the
more robust choice for absolute disorder. Considering only the OST
systems, $H_{odd}^{\alpha_{0}}$ achieves the highest state transfer
fidelities.

For all the channels with $F\rightarrow1$ in the vanishing perturbation
strength limit we find a power law $N\varepsilon_{J}^{\beta}=const$
for the contours of constant fidelity, with $\beta$ near 2, generalizing
the fidelity scaling law found for the linear PST system with relative
disorder \cite{DeChiara2005}. This quantifies the sensitivity of
the channels to perturbations as a function of the system size: Increasing
the channel length, the transfer fidelity becomes more sensitive to
the perturbations.

If the transfer speed is important, independent of the kind of disorder,
the faster transfer is achieved by the nonengineered $H^{\alpha_{opt}}$
system, closely followed by the engineered $H^{lin}$ system. The
other systems are significantly slower.

To summarize, we show that in most situations the transmission performance
of boundary-controlled spin chains renders the full engineering of
the couplings of a spin chain unnecessary in order to obtain quantum
state transmission with high fidelity under static perturbations. 
\begin{acknowledgments}
A.Z. and O.O. acknowledge support from SECYT-UNC and CONICET and are
thankful for the hospitality of Fakultät Physik of the TU Dortmund.
A.Z. is thankful for the support of DAAD. We acknowledge useful discussions
with Dieter Suter.\end{acknowledgments}

\end{document}